\documentstyle[psfig,epsfig]{basi}
\begin{document}
\title[Low Frequency Observations of a Head-Tail
       Radio Source]{Low Frequency Observations of a Head-Tail
       Radio Source}
\author[D.V. Lal \& A.P. Rao]%
       {Dharam Vir Lal\thanks{e-mail: dharam@ncra.tifr.res.in}
       \& A. Pramesh Rao\thanks{e-mail: pramesh@ncra.tifr.res.in} \\ 
        National Centre for Radio Astrophysics
        (TIFR), Pune Univ. Campus, Ganeshkhind,
        Pune 411 007}
\maketitle
\label{firstpage}
\begin{abstract}
We have mapped the head-tail radio galaxy,
3C~129 at 240 and 610~MHz using the GMRT
and studied the detailed morphology and
spectral index variations in this object.
This is the first attempt to observe a
sample of head-tail sources at low frequencies.
We find weak spectral steepening as we go away
from the head along the jet. The Crosspiece has
a spectral index of 0.7
(S$_{\nu} \propto \nu^{-\alpha}$) and is flatter
than the spectral index estimated by Lane et~al.
(2002). We also see a low surface brightness,
diffuse feature at 240~MHz and could be a possible
radio relic candidate (see Figure).
\end{abstract}


`Head-tail' or tadpole shaped radio sources discovered by
Ryle \& Windram (1969, MNRAS, 138, 1) mostly occur in clusters
of galaxies and are characterised by a head identified with
the optical galaxy and two trails sweeping back from the head
(Miley et~al. 1972, Nature, 237, 269).

The VLA map of the strong, prototype head-tail source 3C~129
at 325~MHz shows a small perpendicular object, referred to
as the Crosspiece, near the head of the galaxy. This
feature has a steep spectrum and has been interpreted
by Lane~et~al.~(2002, AJ, 123, 2985) as a pre-existing
fossil radio source that is revived due to 3C~129
ploughing through it. The relic source is compressed
by the bow shock of 3C~129 causing it to radiate behind
the shock front, producing the characteristic shape. If
the above interpretation is correct, structures like the
Crosspiece should be rare. This can be verified by making
high resolution low frequency maps of a well defined
sample  of head-tail sources. With such observations,
we would be able answer, if all head-tail sources,
in general, show such a distinct feature, Crosspiece
{\it or} ~it is characteristic only to 3C~129.

\begin{figure}
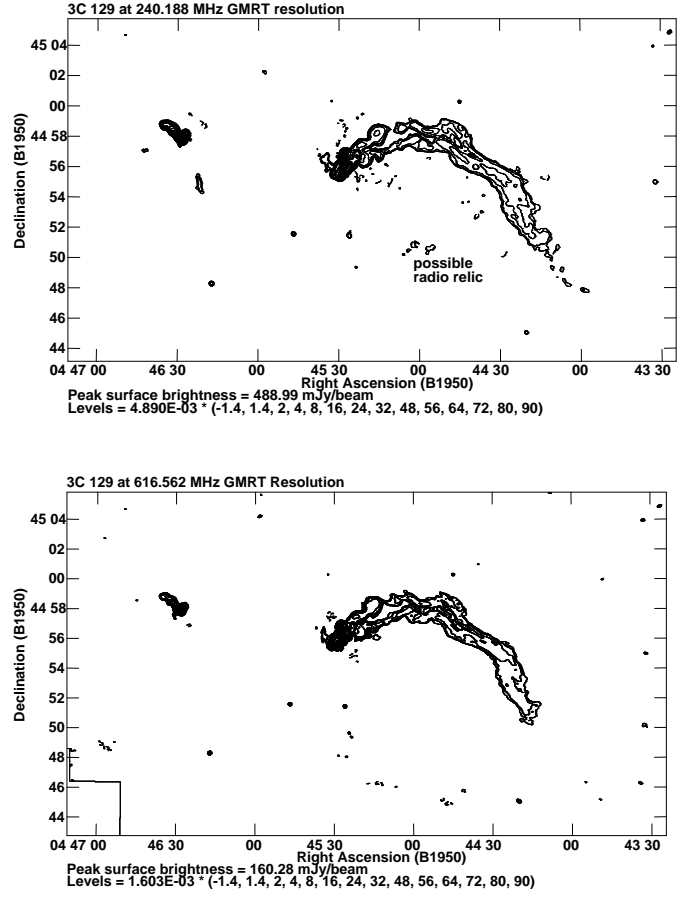

\begin{center}
\begin{tabular}{l}
\includegraphics[width=6cm,angle=270]{240_CONT.PS} \\
\includegraphics[width=6cm,angle=270]{610_CONT.PS}
\end{tabular}
\end{center}
\caption{{\small GMRT images of 3C~129 source at 240 \& 610~MHz.
The possible radio relic source seen in 240~MHz map is shown.}}
\label{blockrcr}
\end{figure}

\section*{Acknowledgements}
\vspace*{-0.2cm}
We thank the staff of the GMRT who have made these observations
possible. GMRT is run by the National Centre for Radio Astrophysics
of the Tata Institute of Fundamental Research.

\label{lastpage}
\end{document}